\documentclass[aps,twocolumn,pra,tightenlines,floatfix,showpacs]{revtex4-1}
\usepackage[dvips]{graphicx}
\usepackage[english]{babel}
\usepackage{amsmath}
\usepackage{amssymb}
\usepackage{times}
\def\be{\begin{equation}}
\def\ee{\end{equation}}
\def\bea{\begin{eqnarray}}
\def\eea{\end{eqnarray}}

\newcommand{\ek}{\xi_{\mathbf{k}}}
\newcommand{\Ek}{{E^{}_{\mathbf{k}}}}

\newcommand{\sumk}{{\sum_{\mathbf{k}}}}

\newcommand{\sumq}{\sum_{\mathbf{q}}}

\newcommand{\Omegaq}{\Omega_{\mathbf{q}}}
\newcommand{\Gammaq}{\Gamma_{\mathbf{q}}}

\renewcommand{\text}[1]{{#1}}

\begin{document}

\title{Zero density limit extrapolation of the superfluid transition
  temperature in a unitary atomic Fermi gas on a lattice}

\author{Qijin Chen}

\affiliation{Department of Physics and Zhejiang Institute of Modern
  Physics, Zhejiang University, Hangzhou, Zhejiang 310027, CHINA}

\date{\today}

\begin{abstract}
  The superfluid transition temperature $T_c$ of a unitary Fermi gas
  on a three-dimensional isotropic lattice with an attractive on-site
  interaction is investigated as a function of density $n$, from half
  filling down to $5.0\times 10^{-7}$ per unit cell, using a pairing
  fluctuation theory.
  We show that except at very low densities ($n^{1/3} <0.2$), where
  $T_c/E_F$ is linear in $n^{1/3}$, $T_c/E_F$ exhibits significant
  higher order nonlinear dependence on $n^{1/3}$.  Therefore, linear
  extrapolation using results at intermediate densities such as in
  typical quantum Monte Carlo simulations leads to a significant
  underestimate of the zero density limit of $T_c/E_F$. Our result,
  $T_c/E_F=0.256$, at $n=0$ is subject to reduction from particle-hole
  fluctuations and incoherent single particle self energy corrections.
\end{abstract}


\maketitle


Experimental realization of superfluidity in cold atomic Fermi gases
has given the BCS--Bose-Einstein condensation (BEC) crossover study a
strong boost over the past decade. More importantly, main interests
have been paid to the strongly interacting regime, where the $s$-wave
scattering length $a$ is large. In particular, the unitary limit,
where the scattering length diverges, has become a test point for
theories. As a consequence, the superfluid transition temperature
$T_c$ in a unitary Fermi gas has been under intensive investigation in
recent years.

Apart from calculating $T_c$ directly in the 3D continuum with various
approximations
\cite{NSR,SadeMelo,Milstein,Griffin,Haussmann2,Zwerger,Floerchinger,Yin2009,Malypapers3,Chen1,ourreview,ReviewLTP,StoofPRL2008,Floerchinger2009,Floerchinger2010,Scherer},
one important method is to calculate $T_c$ on a lattice and then
extrapolate to zero density. It has been argued that the zero density
limit is identical to the continuum case. Indeed, this is the approach
used by quantum Monte Carlo (QMC) simulations. For this approach to
work, two conditions have to be met. First, the result obtained from
the simulation at a given density has to be accurate; this requires
that both the lattice size and the particle number have to be large
enough. Second, the densities $n$ at which the simulations are
performed have to be in the asymptotic linear regime of $T_c$ as a
function of $n^{1/3}$.

It is extremely important to investigate this issue, because the
results of QMC have often been taken with high credibility in the cold
atom community, despite the large discrepancies between the results
from different groups (as well as within the same group sometimes),
and the small total fermion number and lattice size used. For example,
using QMC, Troyer and coworkers \cite{TroyerPRL2006,TroyerPRL2008}
reported $T_c/E_F=0.152$, whereas Bulgac \textit{et al.}
\cite{BulgacMC,Bulgac2009} reported $T_c/E_F =0.23$ and $ 0.15$ in
different papers. Using the method of Ref.~\onlinecite{TroyerPRL2006},
Goulko and Wingate \cite{Wingate} found $T_c/E_F = 0.171$. Another
recent result \cite{Akkineni} from QMC gave $T_c/E_F=0.245$. Although
these different results do not seem to be converging,
the above result of Troyer and coworkers
\cite{TroyerPRL2006,TroyerPRL2008} has been widely cited and compared
with recently. The simulations in Ref.~\onlinecite{TroyerPRL2006} were
done for lattice fermions at finite densities and then extrapolated to
zero density.  It is the purpose of the present paper to investigate
how low in density one needs to go so that the simulations are in the 
%
asymptotic linear regime to ensure the accurateness of the zero
density limit extrapolation.

In this paper, we will study the finite density effect on the zero
density limit extrapolation by calculating $T_c$ on a 3D isotropic
lattice with an attractive on-site interaction, $U$, using a pairing
fluctuation theory. This theory has been able to generate theoretical
results in good agreement with experiment
\cite{ReviewLTP,CompExperiment}. To show how the lattice effect
evolves with fermion density, we drop the complication of the
particle-hole channel. Our result reveals that, as the density
approaches zero, $T_c/E_F$ does reach the 3D continuum value. However,
linear extrapolation using data points calculated at intermediate
densities, such as those in Ref.~\cite{TroyerPRL2006}, will lead to a
significant underestimate of $T_c$ for the continuum limit. When
particle-hole channel contributions are properly included
\cite{ParticleHoleChannel}, we expect that the zero density limit will
yield $T_c/E_F=0.217$, as directly calculated in the continuum.

Details of the pairing fluctuation theory can be found in
Ref.~\onlinecite{Chen1} both in the continuum and on a lattice (see
Ref.~\onlinecite{ParticleHoleChannel} for the treatment of the
particle-hole channel effect).  On a lattice, the fermion dispersion
is give by
%
$\ek = 2t(3-\cos k_x -\cos k_y -\cos k_z) -\mu \equiv \epsilon_\mathbf{k} -\mu ,
$
%
where $t$ is the hopping integral, $\epsilon_\mathbf{k}$ is the
kinetic energy, and we have set the lattice constant $a_0$ to
unity. We define Fermi energy for a given density $n$ by the chemical
potential for a non-interacting Fermi gas at zero $T$. In addition, a
contact potential in the continuum now becomes an on-site attractive
interaction $U$. Namely, we are now solving a negative $U$ Hubbard
model. The Lippmann-Schwinger relation reads
%
${m}/{4\pi a \hbar^2} = 1/{U} + \sumk ({1}/{\epsilon_\mathbf{k}}).
$
%
Therefore, the critical coupling strength is given by $U_c=-1/\sumk
(1/\epsilon_\mathbf{k}) = -7.91355t$. Here $m=t/2$ is the effective
fermion mass in the dilute limit.  In what follows, we shall set
$k_B = \hbar = 1$. 

To recapitulate our theory, the fermion self energy comes from two
contributions, associated with the superfluid condensate and finite
momentum pairs, respectively, given by $\Sigma(K) = \Sigma_{sc}(K) +
\Sigma_{pg}(K)\,$, where
%
$  \Sigma_{sc}(K) = -\Delta_{sc}^2 G_0(-K)$ and $ \Sigma_{pg}(K) = \sum_Q
  t_{pg}(Q)G_0(Q-K) ,
$
%
with $\Delta_{sc}$ being the superfluid order
parameter. $\Sigma_{sc}(K)$ vanishes at and above $T_c$. The finite
momentum $T$-matrix
%
$t_{pg}(Q) = U/[1+U\chi(Q)] 
$
%
derives from  summation of ladder diagrams in the particle-particle
channel, with pair momentum $Q$, where the pair
susceptibility 
%
$\chi(Q) = \sum_K G(K)G_0(Q-K)
$
%
involves the feedback of the self energy via the full Green's function
$G(K)$. As usual, we use a four vector notation, $K\equiv (i\omega_l,
\mathbf{k})$, $Q\equiv (i\Omega_n, \mathbf{q})$, $\sum_K \equiv
T\sum_l \sumk$, and $\sum_Q \equiv T\sum_n \sumq$, where $\omega_l$
($\Omega_n$) are the odd (even) Matsubara frequencies.

By the Thouless criterion, the $T_c$ equation, given by
%
$1+U\chi(0)=0,
$
%
now contains the self energy feedback. This is a major difference
between our pairing fluctuation theory and those based on Nozi\'eres
and Schmitt-Rink (NSR) \cite{NSR} or saddle point approximations \cite{SadeMelo}.

After analytical continuation $i\Omega_n \rightarrow \Omega+i0^+$, one
can Taylor expand the (inverse) $T$-matrix as
%
$t_{pg}^{-1}(\Omega, \mathbf{q}) \approx  Z(\Omega -
  \Omega_\mathbf{q} + \mu_{pair} +i\Gammaq),
$
%
and thus extract the pair dispersion 
$\Omega_\mathbf{q} = 2B(3-\cos q_x -\cos
q_y - \cos q_z)$. 
Here the imaginary part $\Gammaq$ can be neglected when pairs become
(meta)stable \cite{Chen1}.

At and below $T_c$, $\mu_{pair}=0$ and $\Sigma_{pg}(K)$ can be
approximated as
%
$\Sigma_{pg}(K) 
= {\Delta_{pg}^2}/{(i\omega_l+\ek)} + \delta \Sigma 
\approx -\Delta_{pg}^2 G_0(-K) 
,
$
%
with the pseudogap parameter $\Delta_{pg}$ defined as
\be
\Delta_{pg}^2 \equiv -\sum_Q t_{pg}(Q) 
 \approx Z^{-1} \sumq b(\Omegaq) \,,
\label{eq:PG}
\ee
where $b(x)$ is the Bose distribution function.

Neglecting the incoherent term $\delta\Sigma$ in $\Sigma_{pg}$, we
arrive at the total self energy $\Sigma(K)$ in the BCS form:
\be
\Sigma(K) \approx -\Delta^2 G_0(-K),
\label{eq:Sigma}
\ee
where the total gap $\Delta$ is determined via
%
$\Delta^2 =\Delta_{sc}^2 +\Delta_{pg}^2 \,.
$
%
Therefore, the Green's function $G(K)$, the quasiparticle dispersion
$\Ek = \sqrt{\ek^2 + \Delta^2}$, and the gap (or $T_c$) equation all
follow the BCS form, \emph{except that the total gap $\Delta$ now
  contains both contributions from the order parameter $\Delta_{sc}$
  and the pseudogap $\Delta_{pg}$.}
Thus the gap equation is given by
\be
1 + U \sumk \frac{1-2f(\Ek)}{2\Ek} = 0\,,
\label{eq:gap}
\ee
where $f(x)$ is the Fermi distribution function.  
%
In addition, the number equation, $n = 2\sum_K
G(K)$, is given by,
\be
n = \sumk \left[ 1 - \frac{\ek}{\Ek}\,[1-2f(\Ek)] \right] \,.
\label{eq:number}
\ee 
%

Equations (\ref{eq:gap}), (\ref{eq:number}), and (\ref{eq:PG}) form a
closed set. For given interaction $U$, they can be used to solve self
consistently for $T_c$ as well as $\Delta$ and $\mu$ at $T_c$. 

\begin{figure}
\centerline{\includegraphics[width=3.2in,clip]{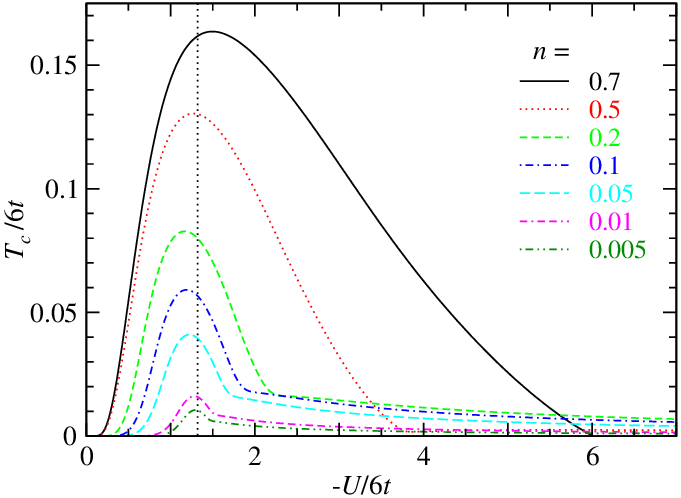}}
\caption{(Color online) Behavior of $T_c/6t$ as a function of the
  attractive on-site interaction $-U/6t$ on a 3D isotropic lattice for
  various density from high to low, as labeled. The unitary limit
  corresponds to $-U/6t = 1.31893$, as indicated by the vertical
  dotted line.}
\label{fig:Tc-U}
\end{figure}

In Fig.~\ref{fig:Tc-U} we plot $T_c$ as a function of pairing strength
$-U/6t$ for various densities from high to low. Here $6t$ is the half
band width. For $n=0.7$, the maximum $T_c$ occurs on the BEC side of
unitarity. Then it moves to the BCS side as $n$ decreases. As $n$
further decreases, the maximum moves slowly back to the unitary point.
This should be contrasted with the 3D continuum case, for which the
maximum occurs slightly on the BEC side. The fact that the maximum
occurs on the BCS side manifests strong lattice effect at these
intermediate densities; it is the lattice effect that causes
difficulty for pair hopping and thus suppresses $T_c$. Even at density
as low as $n=0.005$, the maximum is still slightly on the BCS side.

\begin{figure}
\centerline{\includegraphics[width=3.2in,clip]{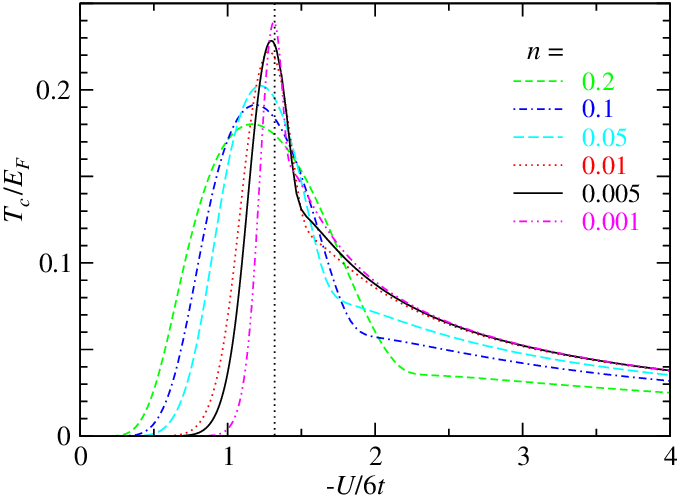}}
\caption{(Color online) $T_c/E_F$ as a function of $-U/6t$ on a 3D
  isotropic lattice for various density from $n=0.2$ to 0.005. The
  unitary limit corresponds to $-U/6t = 1.31893$, as indicated by the
  vertical dotted line.}
\label{fig:TcoverEf-U}
\end{figure}

In order to compare with the continuum $T_c$ curves (see Fig.~10 in
Ref.~\onlinecite{ReviewLTP} for example) more easily, we normalize the
$T_c$ curves by corresponding Fermi energy $E_F$, as shown in
Fig.~\ref{fig:TcoverEf-U}. For clarity, we have dropped the curves for
the two high densities, $n=0.7$ and 0.5. The lattice effect has made
the peak around unitarity much more pronounced, and necessarily
present in all different theoretical treatments of finite temperature
BCS-BEC crossover \cite{NotesonmaximumTc}. As $n$ decreases, this peak
becomes narrower and moves closer to unitarity. Beyond the unitary
limit, the curve for $n=0.001$, as a low density example, exhibits a
rapid falloff with pairing strength, and then decreases following the
functional form $T_c \propto -t^2/U$. This is due to the virtue
ionization during pair hopping in the BEC regime. At unitarity, a
significant fraction of fermions form metastable pairs
\cite{Chen1,ourreview} already at $T_c$, and thus they also see the
lattice effect during pair hopping through virtue ionization. This
suggests that the lattice effect will never go away in the unitary
limit no matter how low the density may be.
Figure~\ref{fig:TcoverEf-U} also reveals that, as $n$ approaches zero,
the maximum $T_c/E_F$ as well as $T_c/E_F$ at unitarity gradually
increase.

\begin{figure}
\centerline{\includegraphics[width=3.2in,clip]{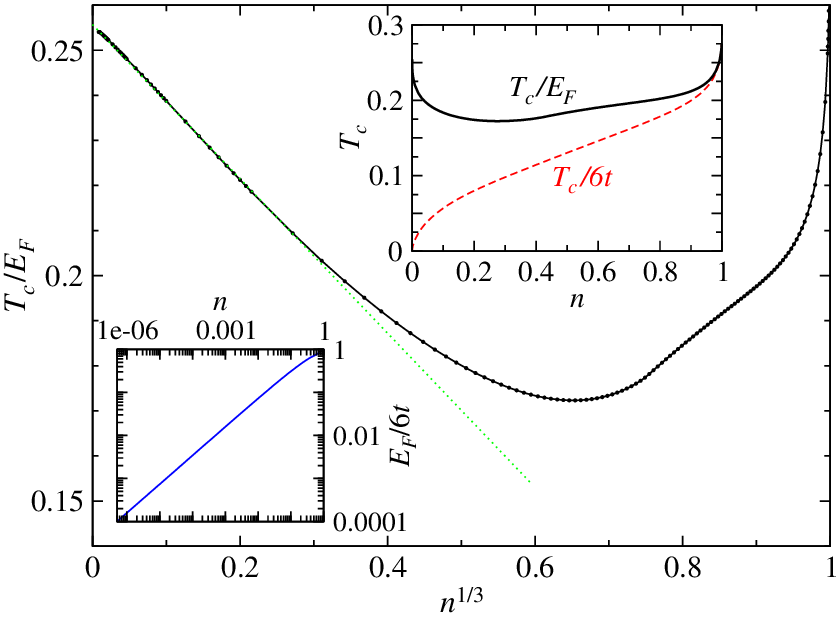}}
\caption{(Color online) $T_c/E_F$ as a function of $n^{1/3}$ on a 3D
  isotropic lattice at unitarity. Shown in the lower left inset is
  $E_F/6t$ as a function of $n$, and plotted in the upper inset are
  $T_c/E_F$ (black solid curve) and $T_c/6t$ (red dashed line) as a
  function of $n$. The (green dotted) linear extrapolation line
  obtained from fitting using data points below $n^{1/3} =0.2$ yields
  $T_c/E_F = 0.2557$ at $n=0$.}
\label{fig:Tc-n}
\end{figure}

Finally, presented in the main figure of Fig.~\ref{fig:Tc-n} is
$T_c/E_F$ as a function of (cubic root of) density $n$ in the unitary
limit, down to $n=5.0\times 10^{-7}$, since the lattice effect is
expected to vary as $n^{1/3}$ to the leading order, namely,
$T_c(n)/E_F(n) = T_c(0)/E_F(0) - \alpha a_0 n^{1/3} + o(a_0^2/n^{2/3})
$, $\alpha$ is a proportionality coefficient. Note that $a_0n^{1/3}$
represents the ratio between the lattice period and the mean
interparticle distance. At the same time, $T_c/E_F$ and $T_c/6t$ are
plotted as a function of $n$ in the upper inset. It shows $T_c/E_F$
increases rapidly near the very end of $n=0$. The behavior of $E_F/6t$
is shown in the lower inset, in a log-log plot. In units of $6t$, both
$E_F$ and $T_c$ vanish at $n=0$ and reach a maximum at half filling
\cite{NotesonCDW}.  In particular, $E_F=6t$ at half filling, as
expected.

Our result reveals that as $n$ decreases from half filling, $T_c/E_F$
decreases and reaches a minimum of 0.172 around $n=0.28$, and then
starts to recover slowly. It does not accelerate until the very end of
$n=0$.  The main plot suggests that $T_c/E_F$ eventually does recover
its continuum counterpart value, 0.256, but the curve exhibits a good
linearity only for $n^{1/3} < 0.2$, i.e. $n < 0.008$. At $n=5.0\times
10^{-7}$, we finds $T_c/E_F=0.254$, close to 0.256. Using the data
below $n^{1/3} = 0.2$, our extrapolation (the green dotted line) leads
to $T_c/E_F = 0.2557 \approx 0.256$ for the continuum limit. Note that
the data points for $n^{1/3} > 0.3 $ shows a rather obvious deviation
from the lower $n$ extrapolation line. This implies that the range of
density for extrapolation used in Ref.~\onlinecite{TroyerPRL2006} is
still far from the asymptotic linear regime. In fact, $n^{1/3} > 0.3 $
cannot be regarded as $\ll 1$. Indeed, the recent result of Goulko and
Wingate \cite{Wingate} seems to confirm this point. They pushed their
simulations down to $n^{1/3} \approx 0.23$ (albeit with a big error
bar), and obtained $T_c/E_F = 0.173$ for the zero density limit using
a linear extrapolation. One can also see from their Fig.~7 that,
without this lower density data point, they would have obtained a
lower value for $T_c/E_F$. In addition, their quadratic fit would
yield $T_c/E_F \approx 0.19$. Finally, we note that a closer look of
Fig. 3 of Ref.~\onlinecite{TroyerPRL2006} suggests that the lowest
density point (also with a big error bar) actually already shows that
their curve starts to bend upward, away from the straight
extrapolation line. Although not conclusive, this observation agrees
with the $T_c/E_F$ curve in Fig.~\ref{fig:Tc-n}.

Despite the big difference between our theory and the QMC approach, it
is reasonable to expect that the lattice effect has a rather similar
effect on $T_c/E_F$.  Therefore, we believe that in order to obtain an
accurate value of $T_c/E_F$ in the zero density limit using a linear
extrapolation, one needs to perform QMC down to $n^{1/3} \sim 0.1$
(i.e. $n \sim 1.0\times 10^{-3}$) or lower.

A later paper by Troyer \textit{et al.}  \cite{TroyerPRL2008} claimed
that they confirmed their lattice fermion result by working in the
continuum limit. However, it is likely that the lattice effect was
actually introduced back through their Eqs.~(3) and (4) and the
periodic boundary condition \cite{Noteonerror}. Indeed, this has been
confirmed by Ref.~\cite{Capone2}, which was partly motivated by the
preprint \cite{preprint} of the current paper. Furthermore, the
thermodynamic limit value was obtained from a linear fit of \emph{only
  3 data points}!  In this later paper, they performed simulations at
unitarity down to $n \approx 0.05$ (with $l_0=1$), or $n^{1/3} =
0.37$. Unfortunately, this density was still far from low enough to
enable an accurate extrapolation.  Above this density, the curve in
the main figure of Fig.~\ref{fig:Tc-n} shows significant deviation
from linearity, caused by contributions of order $n^{2/3}$ and higher.
Our result shows that the densities used for the QMC simulations in
Ref.~\cite{TroyerPRL2006} were not low enough to ensure a good linear
zero density limit extrapolation of $T_c/E_F$ as a function of
$n^{1/3}$.

The QMC simulations by Bulgac \textit{et al.}  were also done on a
lattice, and the number of atoms and the lattice sizes were too small
(e.g. only 50-55 atoms on an $8^3$ lattice in
Refs.~\onlinecite{BulgacMC,Bulgac2009}, equivalent to $n=0.1\sim 0.11$
or $n^{1/3} =0.46\sim 0.48$) to study the dilute limit as we have
done here.
In the context of dynamical mean field theory, Privitera et al
\cite{Capone1} studied the importance of nonuniversal finite-density
corrections to the unitary limit and found that ``densities around $n
\simeq 0.05­-0.01$ are not representative of the dilute regime''.

It should be noted that our $E_F$ is the actual Fermi energy in the 3D
lattice, whereas Troyer and coworkers \cite{TroyerPRL2006} simply
defined $E_F = t k_F^2 = t(3\pi n)^{2/3}$.  This definition will
become the true Fermi energy only in the dilute limit.

Finally, it is interesting to note that on the lattice, the
interaction at unitarity, $U_c$, is density independent so that
$U_c/E_F$ will scale to infinity as $n$ approaches 0. In contrast, in
the continuum, a contact potential can be regarded as the cutoff
momentum $k_0\rightarrow \infty$ limit of an $s$-wave interaction,
$U(k)=U\theta(k_0-k)$, which has $U_c= -2\pi^2/mk_0$. Apparently, this
$U_c$ is scaled down to 0 for a contact potential. This dramatic
contrast for $U_c$ between 3D lattice and 3D continuum seems to
suggest that the 3D continuum cannot be simply taken as the zero
density limit of a 3D lattice. Indeed, the fermions are always subject
to the lattice periodicity no matter how low the density is.


Without including the self energy feedback in the $T_c$ equation
\cite{ourreview}, the NSR theory \cite{NSR,Strinati4,Griffin} and the
saddle point approxmation \cite{SadeMelo} predicted $T_c/E_F=
0.22$. Other approaches reported $T_c/E_F \approx 0.26$
\cite{Milstein}, $0.15$ \cite{Haussmann2}, and $0.16$ \cite{Zwerger},
the last of which exhibits \emph{unphysical} non-monotonic
first-order-like behavior in entropy $S(T)$.
Floerchinger \textit{et al.}  \cite{Floerchinger} found
$T_c/E_F=0.264$ even after including particle-hole
fluctuations. 
Within the present theory, we reported $T_c/E_F= 0.256$
\cite{Chen1,ourreview,ReviewLTP}.

Experimentally, the Duke group \cite{ThermoScience}, in collaboration
with Chen \textit{et al.}, found $T_c/E_F = 0.27$ through a
thermodynamic measurement in a unitary $^6$Li gas. Later, they
\cite{Thomas2007,ThomasJLTP} obtained 0.29 and 0.21 by fitting entropy
and specific heat data with different formulas. The latter value was
obtained assuming a specific heat jump at $T_c$, which may not be
justified in the presence of a strong pseudogap at $T_c$ (See, e.g.,
Refs.~\onlinecite{Loram98,Chen4,ReviewLTP}). According to our
calculations, $T_c$ at unitarity in the trap is only slightly higher
than its homogeneous counterpart, 0.272 versus 0.256. Similar small
difference in $T_c$ between trap and homogeneous cases is expected
from other theories as well.  Therefore, these measurements imply that
the homogeneous $T_c/E_F$ is about 0.25 $\sim$ 0.19. Recently, Ku
\textit{et al.} \cite{Zwierlein2011} reported $T_\text{c}/E_\text{F}
\approx 0.167$ for a homogeneous Fermi gas by identifying the
lambda-like transition temperature.

Our result demonstrates that the $n\rightarrow 0$ limit of the lattice
$T_c/E_F$ does approach that calculated directly in the continuum. We
expect this to remain true when the particle-hole channel
contributions are properly included. In that case, we obtain
$T_c/E_F=0.217$ at unitarity \cite{ParticleHoleChannel}, consistent
with some of the above experimental measurements. Finally, we note
that inclusion of the incoherent self energy $\delta\Sigma$ in our
calculations would further reduce the value of $T_c/E_F$
\cite{ParticleHoleChannel}, bringing it closer to the result of Ku
\textit{et al.}  \cite{Zwierlein2011}.


We thank M. Wingate and M. Capone for useful communications. This work
is supported by NSF of China (grant No. 10974173), MOST of China
(grant Nos. 2011CB921300 and 2011CC026897), the Fundamental Research
Funds for the Central Universities of China (Program No. 2010QNA3026),
Changjiang Scholars Program of the Ministry of Education of China,
Qianjiang RenCai Program of Zhejiang Province (No. 2011R10052), and by
Zhejiang University (grant No. 2009QNA3015).


\bibliographystyle{apsrev4-1}

\begin{thebibliography}{37}
\bibitem [{Not({\natexlab{a}})}]{NotesonmaximumTc}%
  \BibitemOpen
  \href@noop {} {}  \bibinfo {note} {Not every theory
  predicts a maximum in $T_c$ vs pairing strength in 3D continuum with a
  contact potential.}\BibitemShut {Stop}%
\bibitem [{Not({\natexlab{b}})}]{NotesonCDW}%
  \BibitemOpen
  \href@noop {} {}  \bibinfo {note} {We did not include
  particle-hole fluctuations and possible competing charge/spin density wave
  ordering near half filling.}\BibitemShut {Stop}%
\end{thebibliography}

%

\end{document}